\documentclass[default,iicol]{sn-jnl}% Default with double column layout

\usepackage{graphicx}%
\usepackage{multirow}%
\usepackage{amsmath,amssymb,amsfonts}%
\usepackage{amsthm}%
\usepackage{mathrsfs}%
\usepackage[title]{appendix}%
\usepackage{xcolor}%
\usepackage{textcomp}%
\usepackage{manyfoot}%
\usepackage{booktabs}%
\usepackage{algorithm}%
\usepackage{algorithmicx}%
\usepackage{algpseudocode}%
\usepackage{listings}%
\usepackage{microtype}
\usepackage{mathtools}
\usepackage[braket, qm]{qcircuit}
\usepackage{physics}
\usepackage{graphicx}
\usepackage{bm}
\usepackage[caption=false]{subfig} %CHECK AGAIN

\begin{document}

\title[Quantum Gaussian Process Regression for Bayesian Optimization]{Quantum Gaussian Process Regression for Bayesian Optimization}

\author[1]{\fnm{Frederic} \sur{Rapp}}\email{frederic.rapp@ipa.fraunhofer.de}
\author[1]{\fnm{Marco} \sur{Roth}}\email{marco.roth@ipa.fraunhofer.de}

\affil[1]{\orgdiv{Cyber Cognitive Intelligence}, \orgname{Fraunhofer Institute for Manufacturing Engineering and Automation (IPA)}, \orgaddress{\street{Nobelstrasse 12}, \city{Stuttgart}, \postcode{70569}, \country{Germany}}}

\abstract{Gaussian process regression is a well-established Bayesian machine learning method. We propose a new approach to Gaussian process regression using quantum kernels based on parameterized quantum circuits. By employing a hardware-efficient feature map and careful regularization of the Gram matrix, we demonstrate that the variance information of the resulting quantum Gaussian process can be preserved. We also show that quantum Gaussian processes can be used as a surrogate model for Bayesian optimization, a task that critically relies on the variance of the surrogate model. To demonstrate the performance of this quantum Bayesian optimization algorithm, we apply it to the hyperparameter optimization of a machine learning model which performs regression on a real-world dataset. We benchmark the quantum Bayesian optimization against its classical counterpart and show that quantum version can match its performance.}




\maketitle
Quantum computers are expected to have a profound impact on numerous areas in science and industry. The ongoing progress of quantum computing hardware \cite{google_photonic,48651, Bruzewicz_2019} is accompanied by intense algorithmic research activities which explore avenues towards achieving a quantum advantage beyond proof-of-principles \cite{Bravyi_2020, Biamonte_2017}.
Quantum machine learning combines quantum computing and machine learning and is often deemed as one of the fields that could benefit from quantum computing early \cite{liu2021rigorous}. While some quantum machine learning methods rely on running quantum versions of linear algebra sub-routines for a speed-up \cite{Harrow_2009,Rebentrost_2014,Zhao_2019}, these methods usually require deep quantum circuits that are beyond the capabilities of currently accessible noisy intermediate-scale quantum (NISQ) hardware \cite{Preskill_2018}.

Recently, quantum kernel methods have received much attention. These methods are appealing because they can be studied using the well established toolbox of classical kernel theory \cite{representer, Schuld_sup_a_kernels}. Furthermore, using a suitable feature map, they can be implemented on available NISQ devices \cite{Havl_ek_2019}. The general idea is to project the data into the Hilbert space of a quantum computer using a quantum feature map. By calculating pair-wise inner products of data points, a kernel matrix can be calculated which can then be used in classical methods such as support vector machines or kernel ridge regression \cite{svm_scjoell, vovk2013kernel}. The expectation is that by encoding the data into a quantum Hilbert space, the feature map can be enriched with non-classical resources that  provide an advantage compared to classical feature maps. This has already been demonstrated for tailored datasets \cite{liu2021rigorous, Huang_2022}.

While quantum versions of kernel machines like the support vector machine \cite{Rebentrost_2014} have been the focus of recent studies, quantum variants of probabilistic kernel methods have not received as much attention. In this work, we use quantum kernels to create \emph{quantum Gaussian processes} (QGP). Gaussian process (GP) models are popular machine learning methods based on Bayesian inference. GPs are specified by a covariance matrix which can be obtained by calculating the Gram matrix of a kernel function for a given dataset. Given their probabilistic nature, GPs have the desirable property of providing a variance for their predictions which allows uncertainty quantification.

\begin{figure*}[!t]
    \centering
    \includegraphics[scale=0.48]{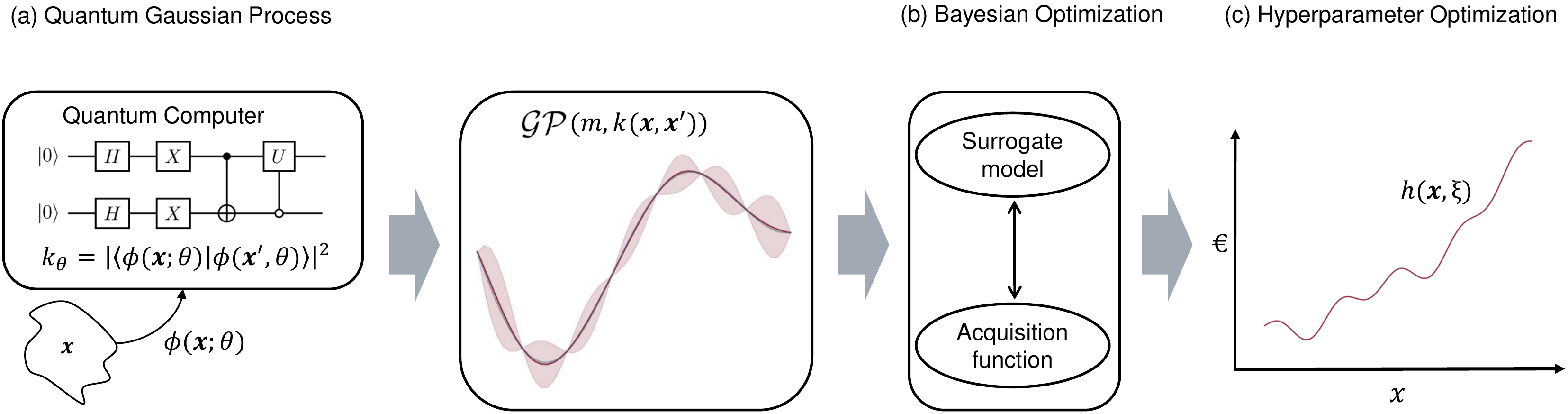}
    \caption{Conceptual layout of the workflow used in this work. (a) The QGP model is constructed by calculating a quantum kernel and substituting the corresponding Gram matrix as covariance matrix into a classical GP. If the feature map used for the quantum kernel contains variational parameters, they can be optimized using maximum likelihood estimation [Eq.~\eqref{eq: NLL}]. (b) By using a QGP model as a surrogate model for Bayesian optimization, a QBO can be obtained. (c) In Sec.~\ref{sec: qbo_results}, the QBO algorithm is used to optimize the hyperparameters $\xi$ of a gradient boosting model $h(\boldsymbol{x},\xi)$ which performs regression on a dataset for remaining value estimation of industrial machines.}
    \label{fig: sketch}
\end{figure*}

Earlier investigation of QGPs have focused on using quantum approximations of classical kernels and have raised the question whether the variance information can be retained in noisy near-term devices \cite{Otten2020QuantumML}. Here, we investigate QGP regression using a hardware-efficient, parameterized feature map. We demonstrate that careful regularization of the Gram matrix can help preserve the variance and show how the overall performance can be improved with an end-to-end optimization using log-likelihood optimization. We show the capabilities of the QGP model by using it as surrogate model for a Bayesian optimization (BO) \cite{10.5555/3379057}, a task that critically relies on the variance information of the surrogate model. We benchmark the resulting quantum Bayesian optimization (QBO) against optimizations using a surrogate models based on conventional GPs and show that QBO can match their performance on the task of optimizing the multidimensional hyperparameters of a classical machine learning model. The hyperparamter optimization is performed on a regression task of a real-world dataset which evaluates the remaining value of used industrial machinery. Figure~\ref{fig: sketch} gives an overview of the various components used in this work.

The manuscript is structured as follows. In Sec.~\ref{sec: qgpr}, we provide an introduction to the fundamentals of QGPs by briefly discussing GP theory and exploring quantum kernels. Subsequently, we illustrate the concept of quantum BO using a QGP surrogate model. In Sec.~\ref{sec: results}, we demonstrate the versatility and effectiveness of QGP models through our analysis of a one-dimensional dataset, followed by their successful application in QBO for the purpose of minimizing a multidimensional function and identifying the optimal hyperparameters of a machine learning model. We present the results of our simulations, including those obtained from noiseless and sample-based experiments, as well as the outcomes from a real quantum computing backend.

\section{Quantum Gaussian Process Regression}
\label{sec: qgpr}
Gaussian process regression is a non-parametric Bayesian machine learning method \cite{Rasmussen_Williams}. It can be used to solve a regression problem of the form 
\begin{equation}
	y = f(\boldsymbol{x}) + \epsilon\,, \label{eq: reg}
\end{equation}
where $f(\boldsymbol{x})$ is a data generating function, with labels $y \in \mathbb{R}$, observed data $\boldsymbol{x} \in \mathcal{X} \subset \mathbb{R}^d$ and independent zero-mean Gaussian noise $\epsilon \sim \mathcal{N}(0,\sigma^2)$. If $f$ is a random function with a Gaussian prior distribution then the function values can be taken as random variables that form a Gaussian process (GP). We denote the GP as $\mathcal{GP}(m,k)$ with a mean function $m$ and a covariance function $k$. Note that $k$ is mathematically equivalent to a kernel function, we will therefore refer to it as \emph{kernel} in the following. A GP is a collection of random variables such that any finite subset is Gaussian distributed \cite{Rasmussen_Williams}. Concretely, for collection of data points $X := (\boldsymbol{x}_1,\dots,\boldsymbol{x}_n)\,,\boldsymbol{x}_i \in \mathcal{X}$ the variables $f(\boldsymbol{x}_i)_{i=1}^n$ are jointly distributed by a multivariate Gaussian distribution such that
\begin{equation*}
	f(\boldsymbol{x}) \sim \mathcal{N}(m(\boldsymbol{x}),k(\boldsymbol{x}, \boldsymbol{x'}))\,,
\end{equation*}
%.
GPs are thus distributions over functions specified by the covariance $k$ \cite{dudley_2002}. 

To predict the values $f_*$ of new data points $X_*$ (test points), we can calculate the posterior distribution given $X$ and $X_*$
% %
% \begin{equation}
% 	\left( {\begin{array}{cc}
% 			f  \\
% 			f_* 
% 	\end{array} } \right) \sim \mathcal{N} \left(0, \left( \begin{array}{cc}
% 		k_{XX} &  k_{XX_*} \\
% 		k_{X X_*}^T &  k_{X_* X_*}
% 	\end{array}     \right)      \right),
% 	\label{eq: fullGram}
% \end{equation}
% %
%
\begin{equation}
	p(f_* \vert X_*, X, f) = \mathcal{N}(f_* ; \mu_*, \Sigma_*)\,. \label{eq: GPR}
\end{equation}
GP regression thus not only yields a prediction for the mean $\mu_*$ but also for the covariance $\Sigma_*$. They are given by
\begin{align}
    &\mu_* = k_{X X_*}^T (k_{XX}+ \sigma^2 \boldsymbol{I})^{-1} f, \label{eq: reg_gp} \\*
	&\Sigma_* = k_{X_* X_*} - k_{X X_*}^T (k_{XX}+ \sigma^2 \boldsymbol{I})^{-1} k_{XX_*}\,.
	\label{eq: noiseGP}
\end{align}
The elements of the Gram matrices $k_{XX}$, $k_{XX_*}$ and $k_{X_* X_*}$ are the pair-wise inner products of the training points, the training and test points, and the test points, respectively. Note that here we have assumed that we only have access to noisy labels as in Eq.~\eqref{eq: reg}. The variance of this noise can be explicitly taken into account in the calculation of the mean and the variance. This servers as an implicit regularization which often results in a better conditioned posterior covariance matrix.

Equations~\eqref{eq: GPR}--\eqref{eq: noiseGP} show that the outcome of the GP is fully governed by the choice of the kernel. In general, a kernel is a positive definite function $k: \chi \times \chi \to \mathbb{R}$, which serves as a similarity measure between pairs of inputs $\boldsymbol{x}$ and $\boldsymbol{x'}$. Specifically, the kernel computes the inner product of the corresponding feature vectors $\phi(\boldsymbol{x})$ and $\phi(\boldsymbol{x'})$ 
\begin{equation}
	k(\boldsymbol{x},\boldsymbol{x'}) = \langle \phi(\boldsymbol{x}),\phi(\boldsymbol{x'}) \rangle_\mathcal{F},\label{eq: classical_kernel}
\end{equation}
in a potentially high-dimensional feature space $\mathcal{F}$, where the feature map $\phi(\boldsymbol{x})$ is a non-linear map from the input space $\chi$ to the feature space $\mathcal{F}$. 

% The well-established mathematical framework of kernel methods makes them a popular choice in machine learning algorithms \cite{Hofmann_2008}.

\subsection{Quantum kernels}
\label{sec: quantum_kernels}

Kernels can be constructed by embedding data into the Hilbert space of a quantum system \cite{Schuld_2019, Havl_ek_2019} [see Fig.~\ref{fig: sketch}(a)]. The resulting quantum state is
\begin{equation}
	\ket{\phi(\boldsymbol{x};\boldsymbol{\theta})} = U(\boldsymbol{x};\boldsymbol{\theta})\ket{0}\,.
	\label{eq: qfm_var}
\end{equation} 
The unitary operator $U(\boldsymbol{x};\boldsymbol{\theta})$ implements the quantum feature map quantum feature map $\phi$. It encodes the classical data point $\boldsymbol{x}$ into a quantum state. In principle, it can depend on additional parameters $\boldsymbol\theta$ that can be trained variationally \cite{Hubreg_}. Using the feature map in Eq.~\eqref{eq: qfm_var}, a quantum kernel can be defined in terms of the Hilbert-Schmidt inner product
\begin{equation}
	k(\boldsymbol{x},\boldsymbol{x'}) = \text{Tr} \left[ \rho(\boldsymbol{x}) \rho(\boldsymbol{x'}) \right],
	\label{eq: q_kernel}
\end{equation}
with the density matrix $\rho(\boldsymbol{x})=U(\boldsymbol{x})\dyad{0}U^\dag(\boldsymbol{x})$. It can be shown that this definition results in a positive definite kernel \cite{Schuld_sup_a_kernels}.
For pure states, Eq.~\eqref{eq: q_kernel} reduces to the overlap between the states encoding the data points such that in practice the kernel elements can be calculated by applying the feature map and its inverse to $\boldsymbol{x}$ and $\boldsymbol{x'}$ and measuring the occupation of the ground state 
\begin{equation}
	k(\boldsymbol{x},\boldsymbol{x'}) = \abs{\braket{\phi(\boldsymbol{x'})}{\phi(\boldsymbol{x})}}^2 = \abs{\bra{0} U (\boldsymbol{x'})^\dagger U(\boldsymbol{x}) \ket{0}}^2.\label{eq: overlap}
\end{equation}
From this it becomes clear that the defining quantity for a quantum kernel $k$ is the quantum feature map $\phi$. The choice of an optimal embedding strategy is an open research question such that the feature maps are often chosen heuristically. Finally, to obtain a quantum GP, we substitute a quantum kernel Eq.~\eqref{eq: q_kernel} into the definition of the variance of a GP model Eq.~\eqref{eq: noiseGP}. This is illustrated in Fig~\ref{fig: sketch}(a).

The variational parameters in Eq.~\eqref{eq: qfm_var} can be trained using various methods. Popular approaches for quantum kernel machines such as  quantum support vector machines or quantum kernel ridge regression often optimize the kernel directly using, e.g., kernel alignment techniques \cite{Hubreg_, kuebler2021inductive, glick2022covariant}. In this work we make use of the Bayesian framework of GPs and train the QGP model end-to-end by maximizing the marginal log-likelihood. Due to the Gaussian form of the posterior [cf. Eq.~\eqref{eq: GPR}], the marginal log-likelihood can be given in closed form \cite{Rasmussen_Williams}
\begin{equation}
\begin{aligned}
	\log p(y\vert X) =& -\frac{1}{2}y^T (k_{XX}(\boldsymbol{\theta})+\sigma^2 I)^{-1} y \\
	&- \frac{1}{2} \log \det(k_{XX}(\boldsymbol{\theta})+\sigma^2 I)\,.
	\label{eq: NLL}
\end{aligned}
\end{equation}
Here, $k_{XX}(\boldsymbol{\theta})$ indicates the dependence of the kernel on the parameters $\boldsymbol{\theta}$ through the parameterized feature map.
The optimization workflow is sketched in Fig.~\ref{fig: sketch}(a). 
Optimizing parameterized quantum circuit is an active area of research with open questions such as how to avoid barren plateaus during training \cite{McClean_2018}. 

In practice, the kernel elements in Eq.~\eqref{eq: overlap} can only be computed approximately because any observable has to be be determined using a finite amount of measurements. The resulting statistical error scales as $\mathcal{O}(1/\sqrt{N})$ where $N$ is the number of measurements. In addition, available NISQ devices suffer from a multitude of noise sources such as short coherence times, gate errors and cross-talk. As a result the estimated kernel $\tilde{k}$ deviates from the true kernel $k$. To ensure that $\tilde{k}$ is positive definite, we need to apply regularization techniques. Taking account of the variance for noisy objective functions as done in Eq.~\eqref{eq: reg_gp}--\eqref{eq: noiseGP} already serves as an inherent regularization. Nevertheless, for noiseless objective functions or for noisy estimates $\tilde{k}$, this might not be sufficient to ensure positive definiteness. Therefore, we employ an eigenvalue-cutoff strategy, where the spectrum of the full Gram matrix 
% %
% \begin{equation}
% k = \left( \begin{array}{cc}
%  		k_{XX} &  k_{XX_*} \\
%  		k_{X X_*}^T &  k_{X_* X_*}
%  	\end{array}     \right)\label{eq: full_gram}
% \end{equation}
% %
is truncated at zero \cite{NIPS1998_7bd28f15}. This requires a full eigenvalue decomposition of the Gram matrix followed by a reconstruction using the truncated spectrum and the original eigenvectors \cite{Hubreg_}. This technique has already been shown to provide good results \cite{Wang_2021}. Additionally, compared to other methods such as shifting the spectrum by the lowest eigenvalue, the truncation does not introduce a constant offset to the variance of the GP model which is desirable for applications where the quantification of uncertainty is required. 
In general, the regularization of Gram matrices used for GP regression is problem-specific and non-trivial, even for classical kernels \cite{reg_comp}.

In this work, we are interested in using QGP models as surrogate models in Bayesian optimization. This is explained in the next section and illustrated in Fig.~\ref{fig: sketch}(b).

\subsection{Quantum Bayesian Optimization}
\label{sec: BayesOpt}
\begin{figure*}[!t]
	\centering
	\includegraphics[scale=0.8]{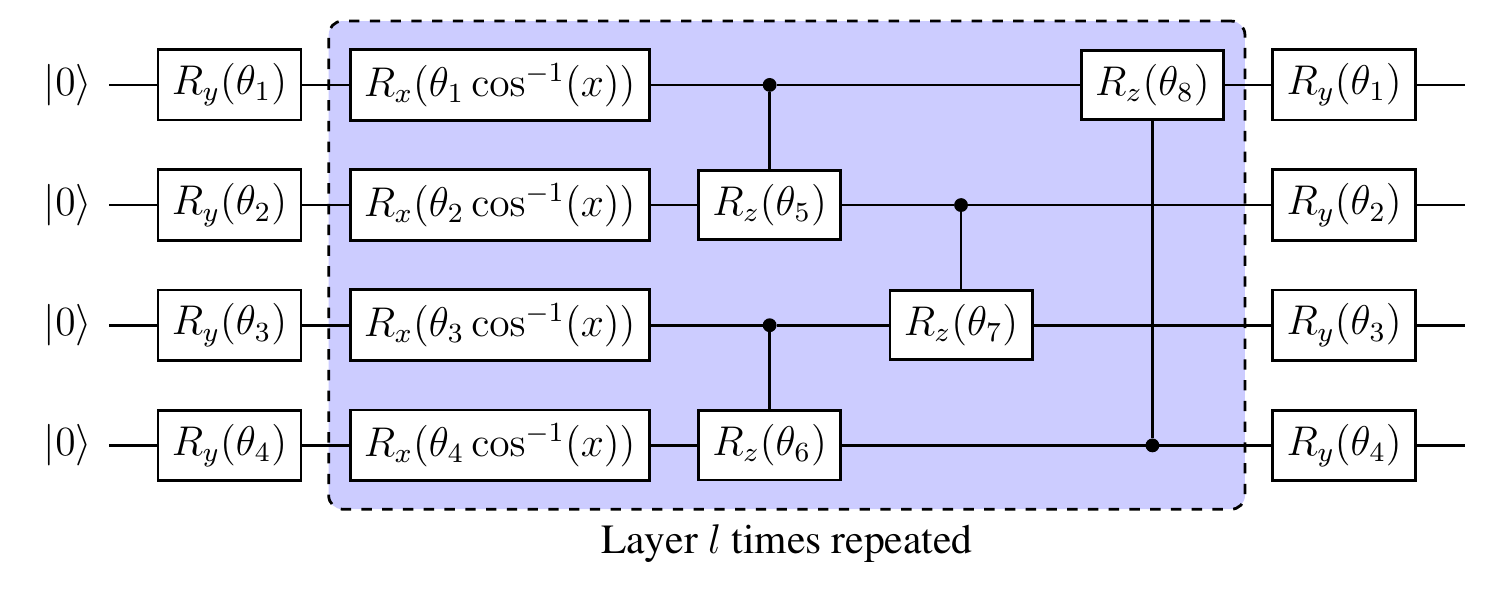}
	\caption{Example of the hardware efficient feature map with $q=4$ qubits and $l=1$ layers, inspired by a Chebychev quantum feature map design~\cite{PhysRevA.103.052416}. The trainable parameters are denoted by $\theta_i$ and the data points by $x$. For the results in this work, various values of $q$ and $l$ are used.
 }
	\label{fig: cheb_map}
\end{figure*}
Bayesian optimization \cite{garnett_2023} is a global optimization method that solves problems of the form
\begin{equation}
	\boldsymbol{x}^* = \text{arg} \min_{\boldsymbol{x}} g(\boldsymbol{x})\,.
	\label{eq :minfx}
\end{equation}
The optimization is performed iteratively where the next sample is chosen using information obtained from previous iterations. Through this informed guidance, BO usually requires a modest amount of samples which makes it attractive for problems where the evaluation of $g$ is expensive. BO treats $g$ as a black-box such that there are no further restrictions regarding its functional form.

The algorithm is initialized by drawing a random sample and fitting a \emph{surrogate model} as a proxy for $g$. The next sample is then chosen by considering an exploitation-exploration trade-off which is quantified by an \emph{acquisition function}. This procedure is then repeated such that the surrogate model approximates the true function increasingly well. Due to their posterior variance output GP models are popular choices for surrogates. A common choice for an acquisition function is the expected improvement (EI) \cite{10.5555/3379057} which measures the expectation of the improvement on the objective $g(\boldsymbol{x})$ with respect to the predictive distribution of the surrogate model. The EI function is given by 
\begin{equation}
	\mathrm{EI}(\boldsymbol{x}) = 
		[g(\boldsymbol{x}^+) - \mu(\boldsymbol{x}) - \lambda] \boldsymbol{\Phi}(Z) + \Sigma(\boldsymbol{x}) \varphi(Z)\,,
	\label{eq: EI}
\end{equation}
 and $\mathrm{EI}=0$ for $\Sigma(\boldsymbol{x}) = 0$. Here $\mu(\boldsymbol{x})$ and $\Sigma(\boldsymbol{x})$ are the posterior mean prediction and the prediction uncertainty of the surrogate model at position $\boldsymbol{x}$, and $\varphi(Z)$, and $\boldsymbol{\Phi}(Z)$ are the probability distribution and the cumulative distribution of the standard normal distribution. The location of the best sample, i.e., the current observed minimum of the surrogate model, is indicated by $\boldsymbol{x}^+$. The standardized prediction error $Z$ is given by $Z = [f(\boldsymbol{x}^+) - \mu(\boldsymbol{x}) - \lambda]/\Sigma(\boldsymbol{x})$ if $\Sigma(\boldsymbol{x}) > 0$ and $Z=0$ if $\Sigma(\boldsymbol{x}) = 0$. The parameter $\lambda$ in Eq.~\eqref{eq: EI} is a hyperparameter that controls the exploitation-exploration trade-off, where a high value of $\lambda$ favours exploration.

We obtain a quantum Bayesian optimization (QBO) algorithm by using a QGP model as a surrogate model. This has the potential to enhance BO for scenarios where quantum kernels have an advantage over classical kernels. A possible drawback is that the exploitation-exploration trade-off, which depends on the model variance is now influenced by quantum computing noise sources. To demonstrate the QBOs capabilities, we apply it to several test cases which is shown in the next section.

\section{Results}
\label{sec: results}

We illustrate the capabilities of QGP models on a one dimensional regression problem. We then demonstrate the feasibility of using QBO with a QGP surrogate model on two multidimensional optimization tasks.
The quantum circuits for the QGP models are implemented using Qiskit \cite{QiskitCommunity2017}. The linear systems for the GPs are solved using a Cholesky decomposition of the Gram matrices.
We validate the algorithm using numerical simulators provided by Qiskit. Results from real quantum computers are obtained from \textit{ibmq{\_}montreal} \cite{ibm_quantum}.
\begin{figure*}[t]
        %\hspace*{-2.0cm}
        \begin{minipage}[t]{.2\textwidth}
            \includegraphics[]{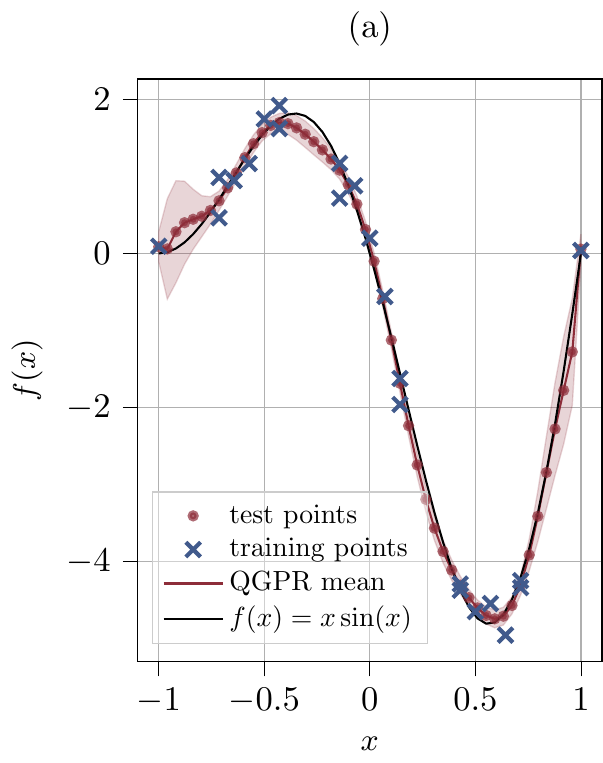}
	\end{minipage}
	%\hfill
        \hspace*{2.8cm}
        \begin{minipage}[t]{.2\textwidth}
            \includegraphics[]{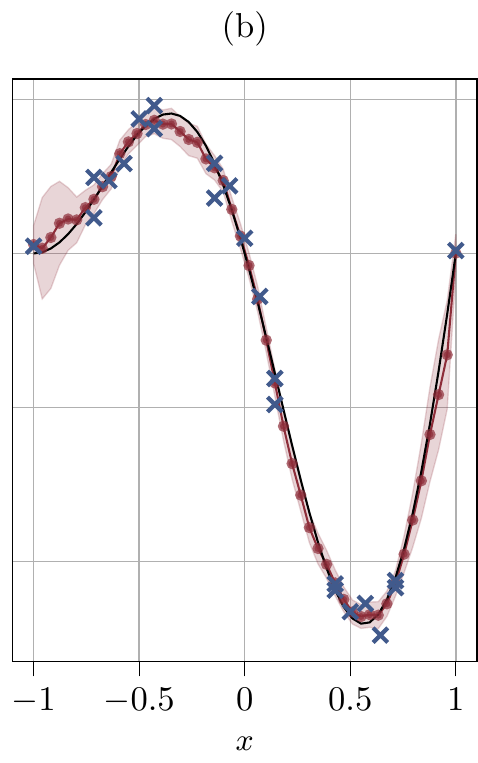}
	\end{minipage}
	%\hfill
        \hspace*{1.5cm}
	\begin{minipage}[t]{.2\textwidth}
            \includegraphics[]{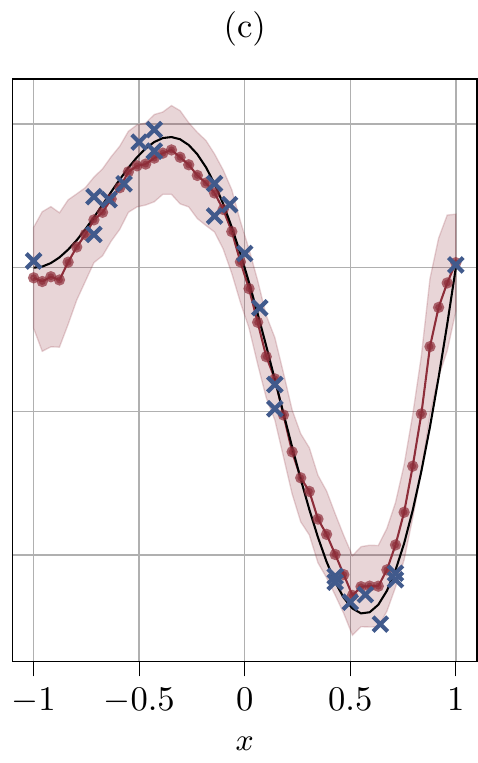}
	\end{minipage}
%        \hfill
        %\hspace*{-4.5cm}
	%\begin{minipage}[t]{.2\textwidth}
	%	\input{Tikz/nll_loss.tikz}
	%\end{minipage}
        %\hspace*{-1.9cm}
	\caption{QGP regression on a dataset created using Eq.~\eqref{eq: sin} (black line). The results are obtained using the feature map in Fig.~\ref{fig: cheb_map} with $q=4$ qubits and $l=2$ layers for the encoding, $n_{\text{training}} = 23$ training points, shown as the blue crosses. The test points are marked by the red dots. The posterior mean of the QGP is shown as the red-line and the standard deviation as the shaded area. (a) shows the result of the statevector simulation with optimized parameters, obtaining an $R^2$ score of $0.996$ and an $ \text{MSE} = 0.022 $. (b) shows the result of the sample-based simulation. We use the optimal parameters obtained in the previous ideal run, resulting in an $R^2$ score of $0.996$ and an $ \text{MSE} = 0.024 $. 
    (c) shows the result of the real hardware run, using the \textit{ibmq{\_}montreal} backend, leading to an $R^2$ score of $0.978$ and an $ \text{MSE} = 0.114 $.
    All runs use the same parameters.}
	\label{fig: cheb_statevector}
\end{figure*}

\subsection{Quantum Gaussian Process Regression}

We apply QGP regression on a one dimensional dataset where the data generating function [cf. Eq.\eqref{eq: reg}] is
\begin{equation}
	f(x) = x \sin(x)\,.
	\label{eq: sin}
\end{equation}
We assume that only noisy labels $y$ can be observed with zero-mean Gaussian noise with a variance $\sigma^2= (0.1)^2$ [cf. Eq.~\eqref{eq: reg}] . We sample $n_{\text{training}} = 23$ non-equidistant training-points in the interval $[0,2\pi]$, and $n_{\text{test}} = 50$ equidistantly-spaced test points. 

The quantum kernel is calculated using a hardware-efficient feature map with variational parameters $\boldsymbol{\theta}$ as depicted in Fig.~\ref{fig: cheb_map}.\footnote{to be published.} We encode the data using $q=4$ qubits and $l=2$ layers. To account for the limited domain of the non-linearity in the feature map, the labels $y$ are scaled to the interval $[-1, 1]$.

To gauge the performance of the model under ideal conditions, we perform statevector simulations from which we obtain completely noiseless quantum kernels. The regression result  can be seen in Fig.~\ref{fig: cheb_statevector}(a) where the mean prediction of the model is shown as a solid line and the standard deviation is depicted as a shaded area. Overall the method is able to achieve a good fit a is visible in the figure. The standard deviation that is obtained from the QGP variance has a reasonable behaviour and is low in areas with high training point density and high in ares where training points are lacking.

\begin{figure}[t]
    \includegraphics[]{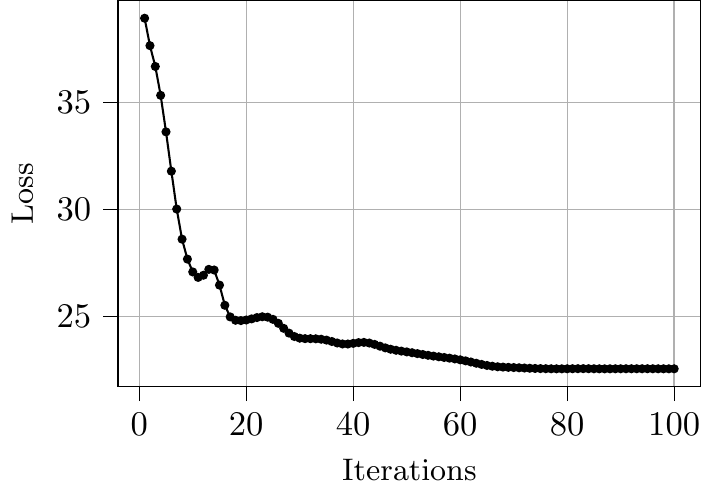}
    \caption{Convergence plot of the log-likelihood loss function [cf. Eq.~\eqref{eq: NLL}], the loss is entirely evaluated on the training data. The variable parameter of the optimization are the angles $\boldsymbol{\theta}$ in the feature map.}
    \label{fig: nll}
\end{figure}

Although good results can already be achieved using a general feature map, e.g., by choosing the parameters $\boldsymbol{\theta}$ randomly \cite{Haug_2023, Jerbi_2023}, we adapt the kernel to the dataset using maximum-likelihood optimization (cf. Eq.~\eqref{eq: NLL} and surrounding discussion). The marginal log-likelihood as a function of optimization iterations can be seen in Fig.~\ref{fig: nll}. In this example, the optimization leads to a reduction of the mean squared error (MSE) by about an order of magnitude [from $0.3$ ($R^2=0.939$) to $0.02$ ($R^2=0.996$)] We observe a convergence of the marginal log-likelihood after $\sim 80$ iterations. The specific optimization behavior is dependent on the chosen feature map design such as the number of qubits, layers and variational parameters. We use the optimal parameters obtained from these ideal simulation for subsequent noisy simulations and calculations on real quantum computers.

Any real quantum computation is ultimately affected by statistical errors. Figure~\ref{fig: cheb_statevector}(b) shows results of the same simulation as in Fig~\ref{fig: cheb_statevector}(a) with sample-based estimation of the wavefunctions with a modest amount of $N=10,000$ measurements per evaluation point. These kind of simulations are a good indicator of the future performance of the model in a regime with low hardware noise. Due to the statistical error in this simulation, the kernel is now only a noisy estimate $\tilde{k}$ of the true kernel $k$. As can be seen in the figure, the performance of the model is only slightly worse compared to the ideal simulation ($\text{MSE}=0.024$). Particularly, due to careful regularization of the Gram matrix (cf. Sec.~\ref{sec: quantum_kernels}) the variance information can be retained reasonably well. 

We conclude this example by running the QGP regression on real quantum hardware using the \textit{ibmq{\_}montreal} device. The results are shown in Fig.~\ref{fig: cheb_statevector}(c). We use readout error mitigation \cite{PRXQuantum.2.040326}, and dynamical decoupling \cite{ezzell2022dynamical} to mitigate the hardware errors. 
Compared to the simulations, the performance of the model slightly decreases with the method obtaining an erorr of $\text{MSE} = 0.114$ on the test data.
Nevertheless, the mean prediction only marginally deviates from the true function. As expected, the regularization of the quantum kernel matrices has to be increased such that the overall standard deviation increases. Nevertheless, even on the real quantum computer the variance of the standard deviation of the prediction can still be retained such that one can clearly distinguish between areas of high and low uncertainty. This is a substantial improvement compared to previous results \cite{Otten2020QuantumML}.

The quality of the solution and the posterior variance are dependent on the chosen quantum feature map. Appendix~\ref{appendix_a} shows results for the same dataset using a different feature map and a different quantum computer.

\subsection{Quantum Bayesian Optimization}
\label{sec: qbo_results}
\begin{figure*}[t]
	%\centering
        \begin{minipage}[t]{.3\textwidth}
            \includegraphics[]{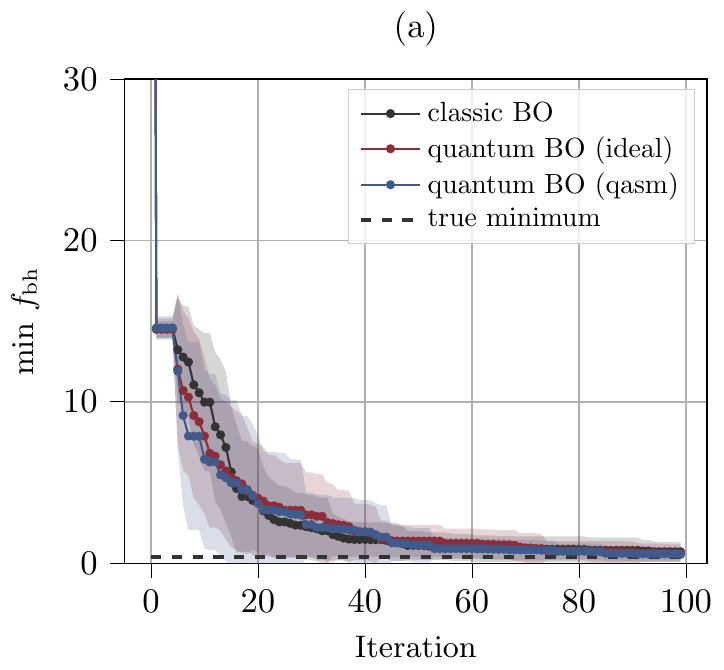}
	\end{minipage}
        %\hfill
        \hspace{3.0cm}
        \begin{minipage}[t]{.3\textwidth}
            \includegraphics[]{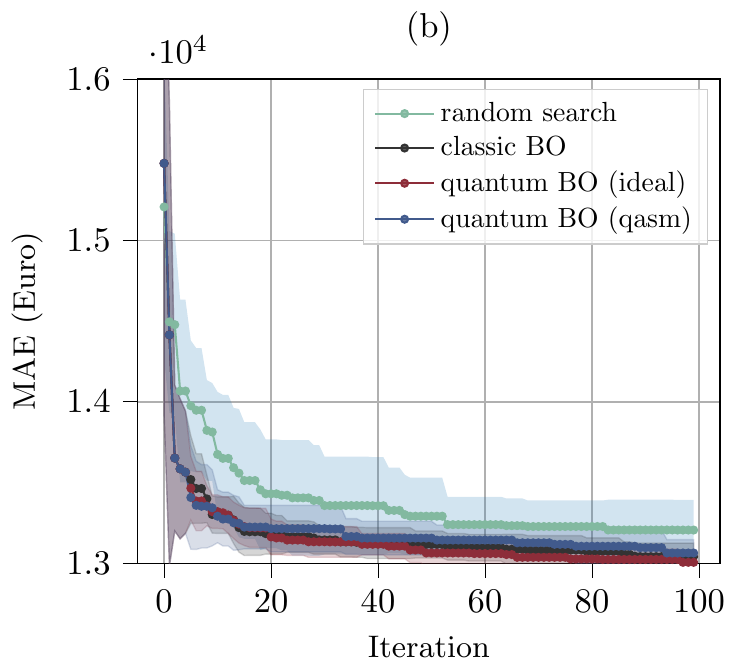}
	\end{minipage}
	
	\caption{BO results averaged over independent runs with the mean shown as solid lines and the variance as shades. The expected improvement [Eq.~\eqref{eq: EI}] is used as acquisition function with an exploration-exploitation parameter of $\lambda = 0.1$ The classical BO uses a GP surrogate model with an optimized RBF kernel (black line). The QBO results are obtained with the feature map in Fig.~\ref{fig: cheb_map} using statevector (red line) and sample-based simulations (blue line). The the initial samples for each individual run are the same for the quantum and classical QBO for better comparison. At each iteration, only the best current result is shown. (a) shows the result for the minimization Eq.~\eqref{eq: branin} where the parameters are fixed at $a=1\text{, }b=5.1/(4\pi^2)\text{, }c = 5/\pi \text{, }r=6\text{, }s=10 \text{ and } t=1/8\pi$. The feature map for the QBO uses $q=4$ qubits and $l=2$ layers. The results are averaged over $25$ runs.
    (b) shows the result of the hyperparameter optimization of a gradient boosting model on a industrial dataset. The average result of ten iterations of random search runs is shown (green, solid). The kernel is calculated using $q=10$ qubits and $l=2$ layers.}
    \label{fig: bopt}
\end{figure*}

We asses the QBO routine introduced in Sec.~\ref{sec: BayesOpt} by minimizing the two-dimensional Branin-Hoo function
\begin{equation}
	f_{\text{bh}}(x) = a(x_2 -bx_1^2 + cx_1 - r)^2 + s(1-t) \cos(x_1) + s\,,
	\label{eq: branin}
\end{equation}
where $a,b,c,s,t$ are real parameters and $x_1 \in [-5,10]\text{, } x_2 \in [0,15]$. We fix the parameters such that the function has three global minima (cf. caption of Fig.~\ref{fig: bopt}). We substitute Eq.~\eqref{eq: branin} to into Eq.~\eqref{eq: reg} to generate data with zero mean Gaussian noise with a variance of $\sigma^2 = (0.5)^2$.

The hardware efficient feature map illustrated in Fig.~\ref{fig: cheb_map} is utilized for the QGP model which is used as a surrogate model for the QBO. We encode the two-dimensional input vector with $q=4$ qubits which increases the model's expressibility compared to a single encoding \cite{Schuld_2021}. Every parameter $\theta$ in the feature map is sampled uniformly from the interval $[0,2\pi]$ and kept fixed for the duration of the optimization.

Figure~\ref{fig: bopt}(a) shows the results for statevector (red line) and sample-based simulations (blue line) where the optimization has been averaged over $25$ runs. The resulting standard deviation of the respective simulations is depicted as shaded areas. It can be seen that both, the BO using kernels obtained from the noiseless and the noisy simulations converge to the true minimum of the function. Especially for the sample-based simulation, this requires thoughtful regularization of the quantum Gram matrices. We compare the performance of the QBO routines to a classical BO with a GP using an RBF kernel. The RBF kernel is optimized in each iteration using maximum likelihood estimation. Despite this optimization which is not used by the QBO it can be seen that the classical and the quantum models perform comparably well.

To demonstrate the applicability of QBO to a real-world scenario, we use the algorithm to optimize the  hyperparameters $\xi$ of a gradient boosting model $h(\boldsymbol{x},\xi)$ \cite{Chen_2016} that is applied to a regression task as illustrated in Fig.~\ref{fig: sketch}(c). The gradient boosting model is used to predict the price of industrial machinery with respect to different machine types, specifications, and amount of working hours. In total, the dataset contains $2910$ data points, and the one-hot encoding of the categorical features leads to $65$ features in total. Further details are shown in Appendix~\ref{appendix_b}. For the optimization, we fix the categorical hyperparameters of the gradient boosting model and only optimize the five continuous hyperparameters (cf. Table~\ref{table: hyperparams}). The objective function for the QBO is the cross validated MAE of the gradient boosting model on the training dataset for a given set of hyperparameters. 

We encode the five dimensional hyperparameter vector with the feature map in Fig.~\ref{fig: cheb_map} using $q=10$ qubits and $l=2$ layers. Figure~\ref{fig: bopt}(b) depcits the result for the different BO runs. Additionally, a random search is shown for comparison. As in the previous example, the QBO results are compared to a BO with a classical GP with an optimized RBF kernel. It can be seen that the results of the QBO are on par with the results of the classical BO. This is true for both, the statevector and the sample-based simulations. As expected, all BO approaches outperform the random search on average. 

\begin{table}[t]
	\centering
	\caption{Hyperparameter space of the gradient boosting model.}
	\begin{tabular}{|c|c|c|}
		\hline
		hyperparameter           & minimum & maximum  \\
		\hline
		$\alpha$   & $0$     & $1.0$    \\
		\hline
		$\gamma$           & $0$     & $5.0$    \\
		\hline
		$n_{\text{max-depth}}$       & $1$     & $50$     \\
		\hline
		$n_{\text{estimators}}$    & $1$     & $300$    \\
		\hline
		$n_{\text{min-child-weight}}$ & $1$     & $10$    \\
		\hline
	\end{tabular}
	\label{table: hyperparams}
\end{table}

\section{Discussion}
\label{sec: discussion}

In this study, we apply QGP models to one and multi-dimensional regression problems and show that they can be used as a surrogate model for BO to create a QBO. We demonstrate that QBO can be used to solve real-world hyperparameter optimization problems. Our encoding strategy allows for effectively using the variational parameters of the data embedding circuit as hyperparameters for the quantum kernels. In our simulations, we observe that the posterior variance of the QGP remains intact under the influence of sampling-noise and even for the calculation NISQ devices, although the influence of the various error sources in the latter affect the result. Nevertheless, since the results from the sampling-based simulations can be seen as an upper-bound for future hardware capabilities, the outlook is optimistic.

Although we demonstrate the feasibility of using QBO to optimize hyperparameters of a machine learning model, the potential benefits of employing quantum kernels over classical machine learning methods in tasks using classical data remain uncertain~\cite{Chia_2020}. However, it is reasonable to expect that QBO may provide advantages in problems where quantum data can be leveraged to achieve a quantum advantage~\cite{Huang_2022}. Notably, QBO is potentially well-suited for active learning tasks in expensive molecular simulations, where the evaluation of the potential energy surface is based on quantum mechanics and is computationally expensive~\cite{article_2, 10.1063/1.5017103}. 
% The evaluation of potential quantum advantages, however, is out of scope of this work and further investigation is necessary to fully understand its utility.

The performance of the QGP model remains unexplored in several avenues within this work. For example, the choice of feature map is a crucial aspect and it has been shown that choosing problem-specific feature map with an inductive bias that is tailored to the dataset has various advantages such as improved performance and trainability \cite{kuebler2021inductive, article}. It is also known that using parametrized feature maps require special care when scaling the number of qubits which can lead to exponential concentration \cite{https://doi.org/10.48550/arxiv.2208.11060}.

Moreover, in this work, we use fidelity-based kernels for the QGP. These have an unfavorable quadratic scaling with the size of the dataset as the pair-wise inner product of the data points have to be calculated. An alternative approach would be to use projected quantum kernels as proposed in \cite{Huang_2021} which not only have a linear scaling but also are thought to have beneficial properties when the dimension of the feature space increases significantly. These alternative kernels could easily be integrated in the QGP and analyzed in future studies. 

While the QGP models presented in this work feature a quantum calculation of the kernel, the majority of their operations are performed classically. However, there is potential for increased improvements by creating a \textit{fully quantum} QGP with a quantum kernel and employing HHL-based inversion of the covariance matrix \cite{Harrow_2009, Zhao_2019}. Such an approach could leverage the benefits of both quantum kernels and quantum linear algebra subroutines, which would help overcome today's limitation of GP models which are currently affected by an unfavorable scaling with the size of the dataset.

\backmatter
\bmhead{Acknowledgments}
This work was supported by the German Federal Ministry of Economic Affairs and Climate Action through the project AutoQML. The authors would like to thank Horst Stühler for kindly providing the dataset. We acknowledge the use of IBM Quantum services for this work. The views expressed are those of the authors, and do not reflect the official policy or position of IBM or the IBM Quantum team.

%\bibliography{scibib}

\begin{appendices}
\section{Industrial Dataset}\label{appendix_b}
The dataset used in this work is shown in Tab~\ref{table: dataset}. It descibes the pricing of industrial machinery in Euros, with respect to several features, e.g. model extensions and working hours. The categorical variables get one-hot encoded, and the price prediction is then carried out using a gradient boosting model. 
\begin{table*}[t!]
	\centering
	\caption{Industrial dataset.}
	\begin{tabular}{|c|c|c|c|c|c|}
		\hline
		Model           & Extension & Location & Working Hours & Price & Construction Year  \\
		\hline
		308   & D     & FI & $8973$ & $44900$ & $2011$   \\
		\hline
		966           & E2   & PL & $4183$ & $57943$ & $2016$   \\
		\hline
		M318      & C      & GB & $3175$ & $86000$ & $2017$   \\
		\hline
		D6    & E      & ES & $22406$ & $58000$ & $2010$  \\
		\hline
		... & K      & NL & $23588$ & $43800$ & $2008$  \\
		\hline
	\end{tabular}
	\label{table: dataset}
\end{table*}
%\section*{Appendix}
%\label{appendix_a}
\section{QGP regression results with different feature map}\label{appendix_a}
\begin{figure*}[t]
        %\hspace*{-2.0cm}
        \begin{minipage}[t]{.2\textwidth}
            \includegraphics[]{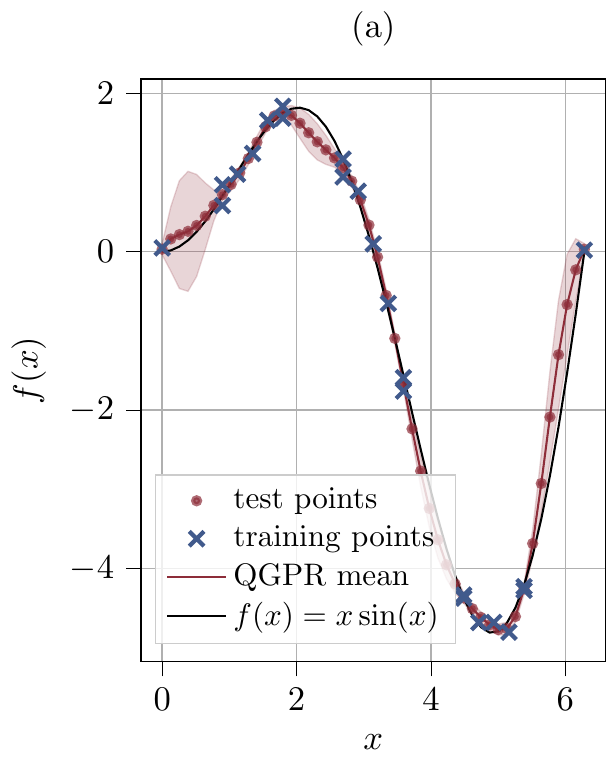}
	\end{minipage}
	%\hfill
        \hspace*{3.0cm}
        \begin{minipage}[t]{.2\textwidth}
            \includegraphics[]{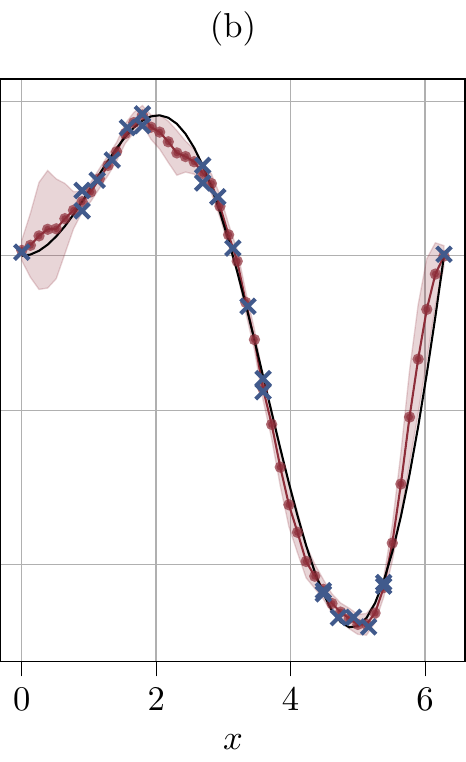}
	\end{minipage}
	%\hfill
        \hspace*{1.5cm}
	\begin{minipage}[t]{.2\textwidth}
            \includegraphics[]{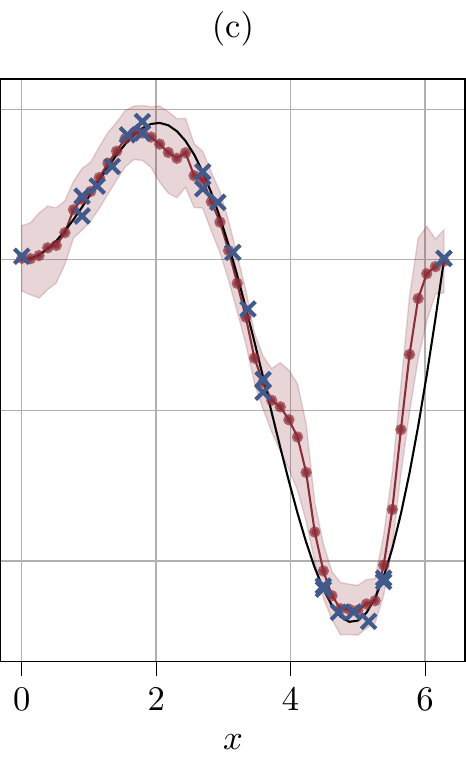}
	\end{minipage}
        %\hfill
        %\hspace*{0.5cm}
	%\begin{minipage}[t]{.2\textwidth}
	%	\input{Tikz/nll_xsinx_statevec_yz_cx.tikz}
	%\end{minipage}
        %\hspace*{-1.9cm}
	\caption{Optimized QGP regression results to regress Eq.~\eqref{eq: sin} (black line). Using $n_{\text{training}} = 23$ training points, shown as the blue crosses, $n_{\text{test}} = 50$ test points, marked by the red dots. The posterior mean of the QGP is shown as the red-line and the standard deviation as the shaded area. (a) shows the result of the ideal simulation with optimized parameters, obtaining an $R^2$ score of $0.986$ and an $ \text{MSE} = 0.07 $. (b) shows the result with sampling noise. We use the optimal parameters obtained in the previous ideal run, resulting in an $R^2$ score of $0.987$ and an $ \text{MSE} = 0.069 $. 
    (c) shows the of the real hardware run, using the \textit{ibmq{\_}ehningen} backend, leading to an $R^2$ score of $0.951$ and an $ \text{MSE} = 0.261 $.
    All runs use the same parameters.}
	\label{fig: ehningen_sim}
\end{figure*}

\begin{figure}[t]
    \includegraphics[]{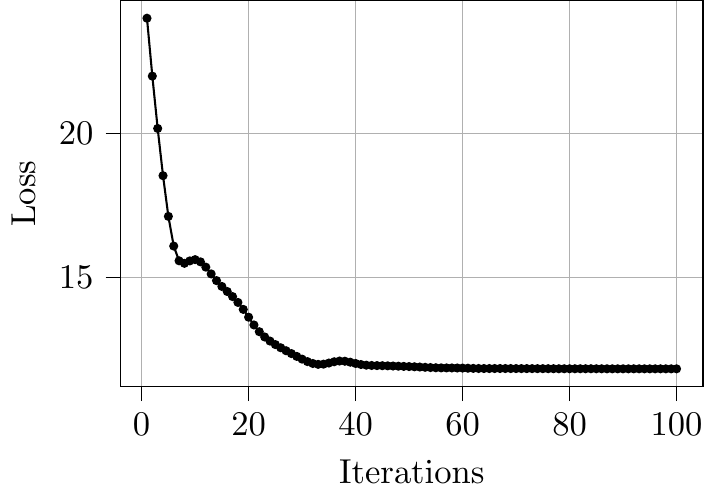}
    \caption{Convergence plot of the log-likelihood loss function [cf. Eq.~\eqref{eq: NLL}], the loss is entirely evaluated on the training data. The variable parameter of the optimization are the angles $\boldsymbol{\theta}$ in the feature map.}
    \label{fig: nll_ehningen}
\end{figure}

% statevector optimization
%\subsection*{QGP regression results with different feature map}
Different choices of feature maps lead to different kernel, and thus to different regression outcomes. Figure~\ref{fig: ehningen_sim} shows the results of the QGP regression as discussed in Sec.~\ref{sec: qgpr} with a different choice of feature map, and on a different quantum computer. 
The feature map here is a hardware efficient design proposed in \cite{Haug_large_scale}. We again use $q=4$ qubits and $l=2$ layers. The real quantum hardware used this time is the \textit{ibmq{\_}ehningen} backend. 
As visible in Figure~\ref{fig: ehningen_sim}(a)-(c) the QGP model is able to regress the objective function very well with the different choice of feature map, and the variance information stays intact. Figure~\ref{fig: nll_ehningen} shows that the loss again converges when optimizing the parameters $\boldsymbol{\theta}$ of the feature map using the ideal statevector simulation.

\end{appendices}
%\begin{figure*}
%\centering 
%\subfloat[]{
%\includegraphics[width=0.3\paperwidth]{trash_example.png}}
%\subfloat[]{
%\includegraphics[width=0.3\paperwidth]{trash_example.png}}
%\subfloat[]{
%\includegraphics[width=0.3\paperwidth]{trash_example.png}}
%\subfloat[]{
%\includegraphics[width=0.4\paperwidth]{trash_example.png}}
%\caption{Hi, I am your caption for this flight.} 
%\end{figure*}

%\input qgp_for_bo.bbl

\end{document}